\documentstyle[twoside]{article}

\newcommand{\refem}{\small\em}
\newcommand{\refbf}{\small\bf}

\catcode`\@=11
\long\def\@makefntext#1{ 
\protect\noindent \hbox to 3.2pt {\hskip-.9pt
$^{{\eightrm\@thefnmark}}$\hfil}#1\hfill} 

\def\thefootnote{\fnsymbol{footnote}}
 \def\@makefnmark{\hbox to 0pt{$^{\@thefnmark}$\hss}}  

\def\ps@myheadings{\let\@mkboth\@gobbletwo
\def\@oddhead{\hbox{} 
\rightmark\hfil\eightrm\thepage}
\def\@oddfoot{}\def\@evenhead{\eightrm\thepage\hfil 
\leftmark\hbox{}}\def\@evenfoot{}
\def\sectionmark##1{}\def\subsectionmark##1{}}

\oddsidemargin=\evensidemargin
\renewcommand{\thefootnote}{\fnsymbol{footnote}}

\newcounter{sectionc}\newcounter{subsectionc}\newcounter{subsubsectionc}
\renewcommand{\section}[1] {\vspace{12pt}\addtocounter{sectionc}{1}
\setcounter{subsectionc}{0}\setcounter{subsubsectionc}{0}\noindent
        {\tenbf\thesectionc. #1}\par\vspace{5pt}}
\renewcommand{\subsection}[1] {\vspace{12pt}\addtocounter{subsectionc}{1}
        \setcounter{subsubsectionc}{0}\noindent
        {\bf\thesectionc.\thesubsectionc. {\kern1pt \bfit #1}}\par\vspace{5pt}}
\renewcommand{\subsubsection}[1] {\vspace{12pt}\addtocounter{subsubsectionc}{1}
        \noindent{\tenrm\thesectionc.\thesubsectionc.\thesubsubsectionc.
        {\kern1pt \tenit #1}}\par\vspace{5pt}}
\newcommand{\nonumsection}[1] {\vspace{12pt}\noindent{\tenbf #1}
        \par\vspace{5pt}}

\newcounter{appendixc}
\newcounter{subappendixc}[appendixc]
\newcounter{subsubappendixc}[subappendixc]
\renewcommand{\thesubappendixc}{\Alph{appendixc}.\arabic{subappendixc}}
\renewcommand{\thesubsubappendixc}
        {\Alph{appendixc}.\arabic{subappendixc}.\arabic{subsubappendixc}}

\renewcommand{\appendix}[1] {\vspace{12pt}
        \refstepcounter{appendixc}
        \setcounter{figure}{0}
        \setcounter{table}{0}
        \setcounter{lemma}{0}
        \setcounter{theorem}{0}
        \setcounter{corollary}{0}
        \setcounter{definition}{0}
        \setcounter{equation}{0}
        \renewcommand{\thefigure}{\Alph{appendixc}.\arabic{figure}}
        \renewcommand{\thetable}{\Alph{appendixc}.\arabic{table}}
        \renewcommand{\theappendixc}{\Alph{appendixc}}
        \renewcommand{\thelemma}{\Alph{appendixc}.\arabic{lemma}}
        \renewcommand{\thetheorem}{\Alph{appendixc}.\arabic{theorem}}
        \renewcommand{\thedefinition}{\Alph{appendixc}.\arabic{definition}}
        \renewcommand{\thecorollary}{\Alph{appendixc}.\arabic{corollary}}
        \renewcommand{\theequation}{\Alph{appendixc}.\arabic{equation}}
        \noindent{\tenbf Appendix \theappendixc #1}\par\vspace{5pt}}
\newcommand{\subappendix}[1] {\vspace{12pt}
        \refstepcounter{subappendixc}
        \noindent{\bf Appendix \thesubappendixc. {\kern1pt \bfit #1}}
        \par\vspace{5pt}}
\newcommand{\subsubappendix}[1] {\vspace{12pt}
        \refstepcounter{subsubappendixc}
        \noindent{\rm Appendix \thesubsubappendixc. {\kern1pt \tenit #1}}
        \par\vspace{5pt}}

\newcommand{\textlineskip}{\baselineskip=13pt}
\newcommand{\smalllineskip}{\baselineskip=10pt}

\def\eightcirc{
\begin{picture}(0,0)
\put(4.4,1.8){\circle{6.5}}
\end{picture}}
\def\eightcopyright{\eightcirc\kern2.7pt\hbox{\eightrm c}}

\newcommand{\publisher}[2]{{\begin{center}\eightrm\smalllineskip
        Received #1\\
        Revised #2
        \end{center}
        }}

\def\abstracts#1#2#3{{
        \centering{\begin{minipage}{4.5in}\baselineskip=10pt\eightrm
        \parindent=0pt #1\par
        \parindent=15pt #2\par
        \parindent=15pt #3
        \end{minipage} }\par}}

\def\keywords#1{{
        \centering{\begin{minipage}{4.5in}\baselineskip=10pt\eightrm
        {\eightit Keywords}\/: #1
        \end{minipage} }\par }}

\renewenvironment{thebibliography}[1]                   
        {\ninerm
         \baselineskip=11pt                             
         \begin{list}{\arabic{enumi}.}
        {\usecounter{enumi}\setlength{\parsep}{0pt}
         \setlength{\leftmargin 17pt}{\rightmargin 0pt} 
         \setlength{\itemsep}{0pt} \settowidth          
        {\labelwidth}{#1.}\sloppy}}{\end{list}}

\newcounter{itemlistc}
\newcounter{romanlistc}
\newcounter{alphlistc}
\newcounter{arabiclistc}

\newcommand{\fcaption}[1]{
        \refstepcounter{figure}
        \setbox\@tempboxa = \hbox{\eightrm Fig.~\thefigure. #1}
        \ifdim \wd\@tempboxa > 5in
           {\begin{center}
        \parbox{5in}{\eightrm \smalllineskip Fig.~\thefigure. #1 }
            \end{center}}
        \else
             {\begin{center}
             {\eightrm Fig.~\thefigure. #1}
              \end{center}}
        \fi}

\newcommand{\tcaption}[1]{
        \refstepcounter{table}
        \setbox\@tempboxa = \hbox{\eightrm Table~\thetable. #1}
        \ifdim \wd\@tempboxa > 5in
           {\begin{center}
        \parbox{5in}{\eightrm\smalllineskip Table~\thetable. #1 }
            \end{center}}
        \else
             {\begin{center}
             {\eightrm Table~\thetable. #1}
              \end{center}}
        \fi}

\def\@citex[#1]#2{\if@filesw\immediate\write\@auxout    
        {\string\citation{#2}}\fi                       
\def\@citea{}\@cite{\@for\@citeb:=#2\do                 
        {\@citea\def\@citea{,}\@ifundefined             
        {b@\@citeb}{{\bf ?}\@warning
        {Citation `\@citeb' on page \thepage \space undefined}}
        {\csname b@\@citeb\endcsname}}}{#1}}

\newif\if@cghi
\def\cite{\@cghitrue\@ifnextchar [{\@tempswatrue
        \@citex}{\@tempswafalse\@citex[]}}
\def\citelow{\@cghifalse\@ifnextchar [{\@tempswatrue
        \@citex}{\@tempswafalse\@citex[]}}
\def\@cite#1#2{{$\null^{#1}$\if@tempswa\typeout
        {IJCGA warning: optional citation argument
        ignored: `#2'} \fi}}

\def\pmb#1{\setbox0=\hbox{#1}
        \kern-.025em\copy0\kern-\wd0
        \kern.05em\copy0\kern-\wd0
        \kern-.025em\raise.0433em\box0}

\def\fnm#1{$^{\mbox{\scriptsize #1}}$}
\def\fnt#1#2{\footnotetext{\kern-.3em
        {$^{\mbox{\scriptsize #1}}$}{#2}}}

\def\runninghead#1#2{\pagestyle{myheadings}
\markboth{{\eightit{\quad #1}}\hfill}{\hfill{\eightit{#2\quad}}}}

\font\tenbf=cmbx10
\font\tenit=cmti10
\font\tenit=cmti10
\font\bfit=cmbxti10 at 10pt
 1
 1
 1

\font\ninerm=cmr9

\font\eightrm=cmr8
\font\eightit=cmti8

\font\tenrm=cmr10

\def\qed{\hbox{${\vcenter{\vbox{                          
   \hrule height 0.4pt\hbox{\vrule width 0.4pt height 6pt
   \kern5pt\vrule width 0.4pt}\hrule height 0.4pt}}}$}}

\newcommand{\bce}{\begin{center}}
\newcommand{\ece}{\end{center}}
\newcommand{\bmath}{\begin{math}}
\newcommand{\emath}{\end{math}}
\newcommand{\bp}{\begin{minipage}[t]}
\newcommand{\ep}{\end{minipage}}

\newcommand{\beq}{\begin{equation}}
\newcommand{\eeq}{\end{equation}}

\def\tmpind#1{\par\noindent\hangindent=0.7truecm\hangafter=1}

\newcommand{\JSP}[1]    {{\refem J.\ Stat.\ Phys.\/} {\refbf #1}}

\newcommand{\JPA}[1]    {{\refem J.\ Phys.\/} {\refbf A#1}}

\newcommand{\PHYSA}[1]  {{\refem Physica\/} {\refbf A#1}}

\newcommand{\NPB}[1]    {{\refem Nucl.\ Phys.\/} {\refbf B#1}}

\newcommand{\PMB}[1]    {{\refem Phil.\ Mag.\/} {\refbf B#1}}
\newcommand{\CP}[1]     {{\refem Contemp.\ Phys.\/} {\refbf #1}}

\newcommand{\NAT}[1]    {{\refem Nature\/} {\refbf #1}}

\runninghead{K. B. Lauritsen, H. Puhl \& H.-J. Tillemans}
{Performance of Random Lattice Algorithms}

\begin{document}

%
%
%
%
%
%
%
%


\thispagestyle{empty}
\setcounter{page}{1}

\renewcommand{\thefootnote}{\fnsymbol{footnote}} 
\def\bsc{{\sc a\kern-6.4pt\sc a\kern-6.4pt\sc a}}
\def\bflatex{\bf L\kern-.30em\raise.3ex\hbox{\bsc}\kern-.14em
T\kern-.1667em\lower.7ex\hbox{E}\kern-.125em X}

\title{\vspace*{-1.7cm}
        \smalllineskip{\flushleft
	%
	%
	%
        {\normalsize HLRZ preprint 30/93}
	\\
        \vspace*{2.1cm}
        {\normalsize\bf PERFORMANCE OF RANDOM LATTICE ALGORITHMS}}}

\author{\footnotesize KENT B\AE{}KGAARD LAURITSEN\thanks{
	\footnotesize Permanent address:
        {\footnotesize\em Institute of Physics and Astronomy, Aarhus
University,
        DK--8000 Aarhus~C, Denmark.\/}
	E-mail: {\footnotesize\tt kent@dfi.aau.dk}},
        ~~HARALD PUHL\thanks{{\footnotesize\em RWTH Aachen, D--52056
        Aachen.\/} E-mail: {\footnotesize\tt h.puhl@kfa-juelich.de}} \\
        {\normalsize and} \\
        {\footnotesize HANS-J\"URGEN TILLEMANS} \\
        {\small\em HLRZ, Forschungszentrum J\"ulich},
        {\small\em D--52425 J\"ulich, Germany}
        }

\date{ }

\maketitle

\vspace{0.10truein}
\publisher{(received date)}{(revised date)}

\vspace*{0.15truein}

\abstracts{
\noindent
We have implemented different algorithms for generating Poissonian
and vectorizable random lattices. The random lattices fulfil
the Voronoi/Delaunay construction.
We measure the performance
of our algorithms for the two types of random lattices
and find that the average computation time
is proportional to the number of points on the lattice.
}{}{}

\vspace*{0.15truein}
\vspace*{0.10truein}


%

\vspace*{1pt}\textlineskip
\section{Introduction}
\vspace*{-0.5pt}

\noindent
There is a large range of applications for random lattices
spanning from materials science,\cite{GHK91}
modelling of the large-scale structure of the
universe,\cite{coles:1990} kinetic growth models,\cite{moukarzel:1992}
sand piles\cite{puhl:1993}
to quantum field theory.\cite{christ-etal:1982}
A random lattice is a way of discretizing a system without
introducing any kind of anisotropy. It can, however, also be used
to describe cellular structures, such as grain mosaics,
biological tissues and
foams.\cite{weaire-rivier:1984,lauritsen-etal:1993,moukarzel:1993,lecaer-ho:1990}
So it is important to have efficient algorithms for generating
random lattices, and, furthermore, algorithms which allow the
random lattice to be dynamically maintained throughout a simulation.

An extensively used type of random lattice is the Delaunay
random lattice,\cite{GS77}
which is a triangulation of space based on a given set
of $N$ points. The \mbox{Voronoi} tessellation\cite{boots-book} is dual to this
lattice and gives a division of space into cells. It is described as
follows: For each point one determines the region of space which is
closer to this point than to any other point.  This procedure called
the \mbox{Voronoi} tessellation of space will divide the space into convex
cells.  There exist other ways of dividing space into cells, such as
the Laguerre partition,\cite{imai-etal:1985}
and, recently, a generalized Voronoi
construction was used in order to model soap froths.\cite{moukarzel:1993}

In the present paper we discuss static and dynamic algorithms for
generating Delaunay random lattices determined according to the
Voronoi tessellation of space.  We will restrict the discussion of the
Voronoi construction to two dimensions; it can be generalized to any
dimension.  We determine the computation time for our algorithms, and
will denote this by $T(N)$ for a random lattice with $N$ points.
Results for both the Poissonian random lattice
(PRL)\cite{HM80} and the vectorizable random lattice
(VRL)\cite{moukarzel:1992} will be given.  The PRL is obtained when the
points are put at independently and randomly chosen positions all over
the entire lattice and then connected according to the Voronoi
construction.  This implies that the PRL random lattice is completely
free of anisotropies.\cite{christ-etal:1982}

The VRL lattice is a modified form of the PRL random
lattice.
Now the space is divided into square cells and each cell contains
exactly one point. The position of a point inside a cell is chosen
randomly.  The motivation for introducing the VRL random lattice was
to have a vectorizable program when solving an equation on the
lattice, e.g., the Laplace equation
describing the numerical
growth of Diffusion-Limited Aggregation (DLA) clusters.\cite{moukarzel:1992}
Furthermore, the VRL lattices possess for all practical purposes
no anisotropies and since a vectorized code can increase
computation time significantly VRL lattices are very useful.

First, in Sec.~2, the Voronoi and Delaunay constructions are described
in more detail.  In Sec.~3 we discuss our different algorithms.
Section~4 contains a discussion of the performances and finally we
conclude.

\section{Voronoi Tessellations and Delaunay Random Lattices}
\label{sec:voronoi}

\noindent
The Voronoi construction or tessellation for a given set of points is
defined as follows: For all points determine the associated cell
consisting of the region of space nearer to this point than to any
other point.  Whenever two cells share an edge they are considered as
neighbours.  By drawing a link between the two points associated
to (located in) the cells one obtains the triangulation of space that is
called the Delaunay lattice. The Delaunay random lattice is dual to
the Voronoi tessellation in the sense that points correspond to cells,
links to edges and triangles to the vertices of the Voronoi
tessellation.

The triangulation (tessellation) of the plane with $N_P$ points (cells),
$N_L$ links (edges)
and $N_T$ triangles (vertices) are constrained
by the Euler relation
\beq
        N_P - N_L + N_T = \chi
\label{eq:euler}
\eeq
where $\chi$ is the Euler characteristic, which equals 2 for a graph on a
surface with the topology of a sphere and 0 for a torus.
Furthermore, $3 N_T = 2 N_L$ holds for a triangulation
(a triangle has three sides/links and $3 N_T$ then counts each link twice).

An equivalent way to obtain the Delaunay random lattice is the
following: Construct for all sets of three
points the circle defined by these points.  If there are no other points
inside this circle the three points will define a Delaunay triangle
and are connected by three links.  The centers of these circumscribing
circles will be the vertices of the
Voronoi tessellation.\cite{christ-etal:1982}

\section{Algorithms}
\label{sec:algorithms}

\noindent
Three different algorithms for generating two-dimensional Delaunay
random lattices and
the associated Voronoi tessellations are presented.  The programs were
written in the C programming language and run on workstations.

First we describe an algorithm for the VRL case, and, secondly, the
PRL random lattice. Thirdly, a dynamic algorithm for generating both the
VRL and PRL lattices is described. We call this algorithm dynamic because
it allows one to change a random lattice slightly and
then obtain the new random lattice by only carrying out local
rearrangements of the links.  Examples of VRL and PRL random lattices
are shown in Fig.~\ref{fig:drl}.

\subsection{Vectorizable random lattices}
\label{subsec:vrl}

\noindent
In the case of the vectorizable random lattice the region where the
points are distributed is first divided into square cells as shown in
Fig.~\ref{vrl_1}(a).  Each cell is of unit area.  Then exactly one
point is put into each cell with the position inside the cell being
random and uniformly distributed.

Each point only can be connected to points in the 36-cell
neighbourhood as shown in Fig.~\ref{vrl_1}(a).  In Fig.~\ref{vrl_1}(b)
the probabilities for the center point to be connected to
another point in this neighbourhood are shown
(cf.\ Ref.~3).
So if a point belongs to a triangle the other
two points are contained in two of the cells of this neighbourhood.
Therefore, to
find a triangle one only has to search for neighbouring points in this
restricted area.
An upper limit for the distance two neighbouring points
can have is $\sqrt{4^2+2^2} = \sqrt{20} \approx 4.472$ as can be seen from
Fig.~\ref{vrl_1}(a).

\begin{figure}[htb]
\centerline{
	}
\centerline{
}
\vspace*{0.2cm}
\fcaption{(a) The division into cells in order to obtain the VRL
	random lattice. Also shown is the initial triangle
	$\scriptstyle ABC$
	together with the 36-cell neighbourhood of the
	initial point $\scriptstyle A$ (see text).
	For the front link $\scriptstyle AB$ a point $\scriptstyle P$
	is searched such that the circle through these three points
        does not contain any other point. The triangle $\scriptstyle APB$ is
        then a Delaunay triangle.
	(b) Probabilities for a point to be connected to another
	point in the 36-cell neighbourhood for a point.}
\label{vrl_1}
\end{figure}

\subsubsection{VRL algorithm}

\noindent
After putting the random points the algorithm determines an initial
triangle (cf.\ Fig.~\ref{vrl_1}(a)).
One point $A$ is chosen. For each pair of points in the neighbourhood
of $A$ it is checked if the circle defined by these three points contains any
other point of the neighbourhood.
The first triplet of points found for which this circle does not contain any
other points is connected
with three links. These links are given a counter clockwise orientation, as
shown by the arrows in Fig.~\ref{vrl_1}(a).

When the lattice is completed
each link is either connected to two triangles or it is a
{\em border link\/}, i.e., it is part of the border which limits the entire
lattice.
Each link is assigned a flag {\tt front} which has the
value 0 if the two triangles connected with this link have been found
or if the link is a border link. If only
one triangle  has been found {\tt front} has the value 1. We call such a link a
{\em front link}, due to the fact that this is the front of the search
algorithm.
So after finding the initial triangle its three links $AB$, $BC$ and
$C\kern-0.05cm A$
are front links. The following steps are now carried out:

For each front link $AB$ check if a point $P$ in the neighbourhood of
$A$ can be
found such that the circumscribing circle of the triangle $APB$
does not contain any other point (cf.\ Fig.~\ref{vrl_1}(a)). This is then
a Delaunay triangle.
First it is checked if the point $P$ is on
the right or the left side of the link $AB$, i.e., if the determinant of
the matrix with the column vectors $AB$ and $AP$ is negative
or positive.\fnm{a}\fnt{a}{
    If $P$ is on the left side,
    $APB$ cannot be a new triangle of the random lattice
    because in this case it has already been found.
    }
Then the next point in the neighbourhood has to be checked.
If no point $P$ completing $AB$ to a Delaunay triangle
can be found the link $AB$ is a border link and its {\tt front} flag is
assigned the
value 0. The program continues by treating the next
front link. If a point $P$ with this property is found there are three new
links $AP$,
$P\kern-0.05cm B$ and
$B\kern-0.05cm A$.
One of them $B\kern-0.05cm A$ is identical with the just checked
front link $AB$, it only has a different orientation. The {\tt front} flag of
$AB$ is assigned the value 0.
The probability to find a connected point to $P$ is not equal
for all cells of the neighbourhood but is much smaller for the outer
cells. We use this by beginning the search at the cells with the highest
probability (cf.\ Fig.~\ref{vrl_1}(b))
and then going outwards to the cells with lower probability of connection.

It is checked if the new links
$AP$ and $P\kern-0.05cm B$ have already been found.
If not, they are new front links. Otherwise they already were saved as front
links and their {\tt front} flag is assigned the value 0.
One notices that it is not necessary to
look through the entire array of links to check whether $AP$ and
$P\kern-0.05cm B$  have already been found.
In our algorithm we use an array {\tt connect} in which we save for each point
the links connected to this point.  Therefore it is only necessary to check
whether the links $AP$ and $P\kern-0.05cm B$
are already saved in the {\tt connect}
arrays of the points $A$ and $B$.
If there is still a front link left treat the next one, otherwise the
Delaunay random lattice is finished.

Finally the Voronoi tessellation is determined. For each point $P$
all triangles which have $P$ as one site are ordered counterclockwise.
The centers of the circumscribing circles
of the triangles are connected. The polygon constructed in this way is the
Voronoi cell which is dual to the point $P$.
In summarized form the algorithm looks like the following:

\begin{itemize}
\item Put random points into the region
\item Construct the initial triangle
\item Construct the Delaunay random lattice. Do the following steps as
      long as there are front links $AB$ left:
  \begin{enumerate}
   \item Find point $P$ which completes front link $AB$ to a Delaunay triangle.
        If not possible, $AB$ is a border link.
  \item Check if the two new links are already stored as front links. If
        not they are new front links.
  \end{enumerate}
\item Construct the Voronoi tessellation by connecting the centers of the
      circumscribing circles of all triangles connected to a point
\end{itemize}

The computation time $T(N)$
for constructing the VRL random lattice is proportional to the number
of points $N$. This is due to the fact that for each point the neighbour points
can only be found in the 36-cell neighbourhood.

\subsection{Poissonian random lattices}
\label{subsec:prl}

\noindent
In this section we describe an algorithm for generating a random
lattice consisting of points with uniformly distributed
$x$ and $y$ values. In the limit of infinite lattices this corresponds
to a Poisson process (see, e.g., Ref.~13). 

The actual generation of a random lattice based on a Poissonian
distribution of lattice points has many aspects in common with the
approach used in the previous Section~3.1. Different
techniques are, however, necessary in some places mainly due to the
fact that for the PRL the number of neighbours is not bounded.  The
algorithm used here is described in detail in Ref.~4 
and based on the one proposed in Ref.~5. 

\subsubsection{PRL algorithm}

\noindent
During the initialization process, a starting triangle is identified
and its edges are defined to be active links, similar to the
front links of the previous section. Thereafter the list of active
links is treated until there are no more active links left. To
identify an active link a flag {\tt treated} is used, which is
initialized to zero for every new link found.  To generate the random
lattice one has then to go through the following steps:

\begin{itemize}
\item Take the next active link:
\begin{enumerate}
\item Draw a circle around this link and check if it contains points
\item If so, choose the one which leads to the largest radius
\item If not, move the center of the circle perpendicular to the link
by a small amount $\epsilon$ dependent on the number of points present
in the region. Draw a
new circle. Repeat until the circle contains points. Select the one
which leads to the smallest radius
\end{enumerate}
\item Store the point found as a new site and create two new links by
connecting it to the two sites of the considered link.
\item Check if these new links already exist. If they do, remove the
active attribute, if not append them as new active links
\item Remove the active attribute from the considered active link
\end{itemize}

\vspace*{0.4cm}
\begin{figure}[htb]
\centerline{
}
%
\vspace*{0.2cm}
\fcaption{Finding a new point to construct a
	new triangle starting from one link of an
	existing triangle. $\scriptstyle M_1$,
	$\scriptstyle M_2$ and
	$\scriptstyle M_3$ are centers of the circles,
	the $\scriptstyle C$'s are points of the lattice.
\label{fig:newpoint}}
\end{figure}

\pagebreak[4]

One has to search for every active link the next
neighbour in the sense of the Delaunay construction. A way to do this
is shown in Fig.~\ref{fig:newpoint}. A circle is drawn through the
endpoints of the link $AB$, its center being the link's midpoint
$M_1$. One has then to check, if there are points inside that circle
(e.g., $C_1$).  If there are, one has to choose the point which gives
the largest radius by drawing a circle determined by the point and the
endpoints of the link. If in the beginning there were no points in the
circle, one has to move the center of the circle by a small amount
$\epsilon$ dependent on the density of points into the direction where
the neighbour should be located (e.g.,~$M_2$, $M_3$), until one or more
points are inside the circle (e.g.,~$C_2$, $C_3$). One chooses then
the point giving raise to the circle with the smallest radius.
This method is used because testing if a point is inside a circle or
not is much faster than the radius calculation.


\subsubsection{Data structure}

\noindent
The two basic elements of the random lattice, sites and links, are
organized as a linked list of special data structures which allow to
store the information needed:

\vspace*{0.3cm}
\hspace*{0.5cm}
\vbox{
\small
\hsize=.9\hsize
\begin{verbatim}
SITE{
  POINT p;              /* Coordinates */
  int border;           /* Border Site Flag */
  int nNb;              /* Number of Neighbours */
  NBNODE *pFirstNbNode; /* Address of the First Neighbour Site */
  SITE *pNextSite;      /* The Next Site in the List */
  SITE *pNextSlSite;    /* The Next Site in the Current Bin */
};
\end{verbatim}
}

\vspace*{0.3cm}
\hspace*{0.5cm}
\vbox{
\small
\hsize=.9\hsize
\begin{verbatim}
LINK{
  POINT p1, p2;         /* Coordinates of Constituting Points */
  SITE *pSite1,*pSite2; /* Addresses of these Points */
  int orient;           /* Orientation of the Link */
  char border;          /* Border  Link Flag */
  char treated;         /* Treated Link Flag */
  LINK *pNextLink;      /* The Next Link in the List */
};
\end{verbatim}
}

\noindent
The site list is initialized only once in the beginning. During the
construction of the lattice neighbours are, however, dynamically
appended to each site. The address of the first neighbour is given by
{\tt pFirstNbNode}.  Each appended neighbour is again a structure
containing information about the address of the neighbour site and
the link connecting it:

\vspace*{0.3cm}
\hspace*{0.5cm}
\vbox{
\hsize=.9\hsize
\small
\begin{verbatim}
NBNODE{
  NBNODE *pNextNbNode;  /* The Next Neighbour Site */
  SITE *pNb;            /* Address of the Current Neighbour */
  LINK *pLink;          /* Address of the Link Connecting */
};                      /* the Site with this Neighbour */
\end{verbatim}
}

In contrast to the arguments given in Section~3.1 one
does not know where exactly to find the neighbour site. For the search
of this neighbour it is however not acceptable to do a complete search
of all points, since this would imply a time behaviour proportional to
$N^2$. Alternatively one could use a tree-like structure, where one
can identify points in the neighbourhood. Another possibility is to
implement binning, i.e., divide the lattice into sublattices so that
in each bin one finds on the average the same number of sites. This
method has been implemented here. The pointer {\tt pNextSlSite}
contains the address of the next site in a considered bin. Thus for
each link---on the average---only a fixed number of sites has to be
searched. This number should be equal to 5--10 sites per bin, while
the $8$ neighbouring bins are searched for each active link.

During the construction of the lattice the algorithm will encounter
situations, where a neighbour found gives raise to new links which
have already been found, and are still waiting in the active link list
to be treated. In this case one has to mark these links as treated by
setting the corresponding flag {\tt treated} in the link structure.
Treated links are skipped when processing the link list.

The stopping criterion depends on the problem one is investigating.
If there are no periodic boundary conditions, one has to introduce a
limit beyond which the algorithm should stop to look for new points.
If the problem has periodic boundary conditions, no special stopping
criterion is needed.

\subsubsection{Speed and memory}

\noindent
The speed of a program depends on the setup of the algorithm and the
realization on a given computer. The best performance one could have
is $O(N)$, i.e., the order of the computation time $T(N)$ increases
linearly with the number of points treated.

To avoid any superflous calculation the following measures have been
taken in this algorithm:
\begin{itemize}
\item While creating the lattice one has to check whenever a new neighbour
and therefore two links are found, that these links are not yet in the
list of active links. To minimize the impact of this task on the time
behaviour, each site knows through the structure {\tt NBNODE} not only
about its neighbours {\tt pNb} but also about the connecting links
{\tt pLink}. So only the links already attached to the site have to be
searched, a number which is constant on the average.
\item Searching potential neighbours, one has to draw circles and move
their centers. Care is taken, that the quantity $\epsilon$ by which
the centers are moved depends on the density of points. If one would
not account for that, more and more heavy radius calculations would be
necessary as the density grows.
\end{itemize}
With these measures taken the algorithm should have a time behaviour
like $O(N)$, since for every point---on the average--a fixed number of
calculations and memory allocations are carried out.

These theoretical considerations do, however, not reflect reality.
Dynamic memory allocation is intuitively and easily programmable; the
computer will on the other hand be obliged to maintain longer and
longer lists of allocated memory blocks, so memory allocation time
will grow faster than the system size.

That is why in this program memory is allocated in huge blocks from
the start. Whenever new memory is needed (for new neighbours or links)
it is taken from these ``pools''. The computer thus has no additional
work to do for maintaining memory allocation tables.\fnm{b}\fnt{b}{
       The total time for computing large lattices will increase more than
       linearly since the computer will be forced to swap memory pages.}
Allocating memory in the beginning
is in fact possible since one knows that the {\em average\/}
number of neighbours for a site is 6. Particularly for large lattices
($N\geq$ 10,000) this is very advantageous, since memory management
tends to become a substantial part of the calculation time. Allocating
memory from the start one has the other advantage that the linking of
the lists is not necessary anymore, and this saves memory.

\subsection{Dynamic Random Lattices}
\label{subsec:drl}

\noindent
Now we describe a method to calculate a random lattice
using a Dynamic Random Lattice algorithm (DRL), which makes it
possible to dynamically maintain the random lattice in a
simulation.\cite{lauritsen-etal:1993,lauritsen-moukarzel:1993}
The idea is to start from an existing lattice (e.g., a triangular lattice)
and then successively change this lattice by moving the points in order
to obtain the final random lattice.
The lattice fulfils at all times the Voronoi
construction and this ensures that also the final lattice will
be a Delaunay random lattice.

\begin{figure}[htb]
\centerline{\vbox{\hbox{
}\hbox{
}}}
\vspace*{0.3cm}
\fcaption{Random lattices, (a) and (b) VRL lattices, (c) and (d) PRL
	lattices, with periodic boundary conditions
	obtained with the DRL algorithm.
	The lattices contain 400 points and were obtained by starting
	from a triangular lattice.}
\label{fig:drl}
\end{figure}

By the DRL algorithm, the VRL random lattice is
obtained by moving the points---one
by one---to new positions randomly chosen in the cells
(cf.\ Fig.~\ref{vrl_1}(a))
associated to the points (see Figs.~\ref{fig:drl}(a)-(b)).
The PRL random lattice is obtained by moving the
points to new positions randomly chosen over the entire lattice
(see Figs.~\ref{fig:drl}(c)-(d)).
The time dependence for the DRL algorithm is more complicated than for the
other two algorithms due to the
``search''  part of the algorithm (see below) and will have the slowest
performance in the case of the PRL random lattice and the fastest
in the case of the VRL random lattice.

\subsubsection{DRL algorithm}

\noindent
We now describe the basic elements of the algorithm
(see Refs.~7 and~14 
for a more detailed description).
When a point is moved from one place to another it is possible to
view all the changes of the lattice as link-flip processes.
A link-flip process is when, e.g., the link $EP$ in Fig.~\ref{fig:flip}(a)
is flipped to $DF$ as in Fig.~\ref{fig:flip}(b).
In this case it is a neighbour-loosing process for the points $P$
and $E$, whereas it is a neighbour-gaining process for the
points $D$ and $F$.
Intuitively, when a point is moved,
one follows the trajectory of the point and tests
whether a link-flip process will take place. The condition for this
to occur is: i) either the point $P$ enters the circumscribing
circle of a triangle
(corresponding to the points $A$, $B$ and $C$ in Fig.~\ref{fig:flip}(a)),
which then implies a neighbour-gaining process (cf.\ Fig.~\ref{fig:flip}(b))
or, ii)~the point $P$ leaves the circumscribing circle for
three points
(corresponding to the points $D$, $E$ and $F$ in Fig.~\ref{fig:flip}(a), which
do not define a triangle),
and this implies a neighbour-loosing process (cf.\ Fig.~\ref{fig:flip}(b)).

\begin{figure}[htb]
\centerline{
}
\vspace*{0.3cm}
\fcaption{Part of a Delaunay random lattice (a) before and (b)
        after the point $\scriptstyle P$ has been moved.
	The link-flip process appearing to
        the left of the point $\scriptstyle P$ is a neighbour-loosing process
        whereas the process to the right is a neighbour-gaining process.
        The thick lines represent the links of the random lattice
        whereas the thin lines are the edges of the Voronoi cells.}
\label{fig:flip}
\end{figure}

Even though it is impossible to predict the number of
processes that will take place, the full set of changes can always be
viewed as consisting of the neighbour-gaining and loosing processes
taking place by following the point along its trajectory.
It is, however, not necessary to carry out the updating
of all the processes along the points trajectory. When a point has
to be moved one first {removes} it from the lattice leaving
a lattice with $N-1$ points, and updates the
neighbourhood. Then the point is put at its new
position, and the location of this new position,
i.e., to which triangle it belongs in the lattice,
is determined.
Since one does not know into which of the existing triangles
on the random lattice the point is put,
the way this triangle is determined is by a {search} through the
lattice (see Ref.~14). 
Finally, the point is {added} and
the neighbourhood is updated by carrying out neighbour-gaining and
loosing processes.
In schematized form the DRL algorithm looks like this:


\begin{itemize}
\item Generate a triangular lattice with $N$ points
	and the topology of a torus (periodic boundary conditions)
\item For all points $P$ on the lattice do:

	\begin{enumerate}
	\item {\em remove\/} the point $P$ and the links to its neighbours
	\item calculate/update the new links
	\item put the point $P$ at a new (randomly chosen) position
	\item {\em search\/} through the Delaunay random lattice with
	      $N-1$ points and locate in which triangle the new point belongs
	\item {\em add\/} the point $P$ and calculate/update the new links
	      followed by
	      neighbour flipping processes (cf.\ Fig.~\ref{fig:flip})
	\end{enumerate}
\end{itemize}

\subsubsection{Data structure}

\noindent
The random lattice algorithm is implemented using a dynamic
pointer structure.
For each point a {\tt SITE} structure (see below)
is defined where information associated to the
site is stored.  Included in this information
is a nearest-neighbour pointer list {\tt NN}. For
each site in this list there is a pointer referring
to the location of the original site itself, so that the information
associated to a site is only stored once.
The sites and nearest neighbour linked lists are organized as follows:

\vspace*{0.3cm}
\hspace*{0.5cm}
\vbox{
\small
\hsize=.9\hsize
\begin{verbatim}
SITE {
  POINT p;     /* coordinates */
  int   n;     /* point number */
  int   q;     /* coordination number */
  NN    *pNN;  /* pointer to first Nearest Neighbour */
};
\end{verbatim}
}

\vspace*{0.3cm}
\hspace*{0.5cm}
\vbox{
\small
\hsize=.9\hsize
\begin{verbatim}
NN {
  SITE   *pSite;    /* pointer to the site */
  CIRCLE circle;    /* circumscribing circle */
  int    pos;       /* position relative to the site */
  NN     *pNN;      /* pointer to the Next Neighbour */
};
\end{verbatim}
}

The updating of the lattice when a flip process takes place is
carried out by rearranging the {\tt pNN} pointers in the nearest-neighbour
pointer lists.
When a new triangle appears or an old one disappears information on
the circumscribing circles is updated and stored in the {\tt NN}
nearest-neighbour pointer list (cf.\
Ref.~14).

\subsubsection{Performance}
\label{sec:drl-perf}

\noindent
The average computation time of the DRL algorithm can be written as
\beq
	T(N) = O \left(N (R(N) + S(N) + A(N) ) \right)
\eeq
where $R(N)$ is the average time to remove a point, $S(N)$ the
search time and $A(N)$ the time to add a point.
For a triangular lattice the coordination number is 6 so
$R(N)$ and $A(N)$ would both imply 6 nearest-neighbour
operations. From Eq.~(\ref{eq:euler}) follows that
the average coordination number for a triangulation is 6,
so both $R(N)$ and $A(N)$ will on average be approximately constant.

The search time $S(N)$ is proportional to the distance between the old and
new position of the point.\cite{lauritsen-moukarzel:1993}
In the VRL case the new positions of the points
are within a fixed distance from the old positions, i.e.,
$S(N)$ is a constant, and we expect the total time $T(N)$ to scale linearily
with $N$, i.e., an $O(N)$ behaviour.
For the PRL lattice the search will in the worst case be
proportional to the size of the system, i.e., $\sqrt{N}$,
leading to an $O(N \sqrt{N})$ behaviour for the computation time.

\section{Results}
\label{sec:results}

\noindent
All simulations we report in this section were carried out using Sun
Sparc 10 workstations.
In Table~\ref{tab:vrl} and Fig.~\ref{fig:time}
are shown the computation times for the static VRL
and PRL algorithms as well as for the dynamic algorithm, denoted here
as DVRL and DPRL.
For the static version one notices an approximate linear increase in
computation time as a function of the number of points $N$ on the
lattice. For large systems the performance is worse due to the fact
that almost all the memory of the computer is used.
In the VRL case the expression \mbox{$T(N) \approx 0.20 N$}
milliseconds fits the
data very well whereas the PRL algorithm is a little slower
with the expression
\mbox{$T(N) \approx 0.24 N$} milliseconds providing a good fit.\fnm{c}\fnt{c}{
	The linear behaviour of the algorithm is only valid if every
	bin contains about the same number of points. For
	distributions of $x$ and $y$ values which are not
	uniform, the binning should be adjusted.
}

\begin{table}[htb]
\tcaption{The upper part shows the performance of the VRL algorithm for
the VRL random lattice and for the PRL algorithm the PRL random
lattice. In the lower part the values for the dynamic algorithm are
given for the VRL and PRL random lattice, denoted here as DVRL and
DPRL. The times are measured in seconds.}
\vspace{0.3cm}
\centerline{
\begin{tabular}{l|rrrrrrr}
	\hline
\# Points, $N$ &  100 & 1000 & 10,000 & 50,000 & 100,000 & 200,000 & 300,000 \\
	\hline\hline
	VRL            & 0.04 & 0.18 &   1.72 &   8.90 & 18.3 & 38.6 & 60.3 \\
	PRL            & 0.02 & 0.20 &   2.05 &   11.3 & 23.0 & 47.3 & 73.9 \\
	\hline
	\hline
	DVRL           & 0.11 & 0.99 &  10.15 &   50.6 & 101.0 &  207 &  308 \\
	DPRL           & 0.14 & 1.70 &  29.20 &  264.8 & 713.8 & 1938 & 3476 \\
	\hline
\end{tabular}
}
\label{tab:vrl}
\end{table}
\begin{figure}[htb]
\centerline{
}
\vspace*{0.3cm}
\fcaption{Computation time for the VRL, PRL and DRL algorithms in order to
	generate random lattices with $\scriptstyle N$ points. The times are
	measured in seconds. Also shown are curves representing a linear
	increase and an $\scriptstyle N \sqrt{N}$ increase in computation time.}
\label{fig:time}
\end{figure}

For the DRL algorithm one notices a linear relation between execution
time and the number of points $N$ for the VRL random lattice,
$T(N) \approx 1.01 N$ milliseconds.
In the PRL case the behaviour asymptotically approaches the
$O(N \sqrt{N})$ behaviour, as discussed
in Sec.~3.3.3.\fnm{d}\fnt{d}{
	The initialization time for generating the initial triangular
	lattice where memory for the dynamic pointer structure is allocated
	is included in the above computation times. This
	initialization time scales
	approximately linearily as $cN$ with $c \approx 0.157$ ms.}
An expression of
the form $T(N) = aN (b + \sqrt{N})$
(with $a \approx$ 150 ms and $b \approx 65$)
follows our curve (cf.\ Fig.~\ref{fig:time}), and this
has the anticipated $O(N \sqrt{N})$ behaviour for large $N$.
When in advance one has a knowledge of the distribution of the points
it may be possible to reduce the search time by using ``bins''
(cf.\ Sec.~3.2.2), but we have only investigated the general DRL
algorithm without using any information about the location of the
points.

\section{Conclusions}

\noindent
In the present paper we have discussed different algorithms for
generating Delaunay random lattices and the associated (dual) Voronoi
tessellations. We have described how the algorithms have been
implemented in the C programming language using dynamic pointer
structures.

For the vectorizable random lattice and the Poissonian random lattices
we found that the computation time is roughly linear. When the system
size approaches the memory limit of the computer the time
dependence increases more rapidly. We also discussed a dynamic
random lattice algorithm. This algorithm allows for a random lattice to
be maintained during a simulation and can also be used in order
to obtain a random lattice but with a slower time performance
than the VRL and PRL algorithms.

\nonumsection{Acknowledgments}

\noindent
We thank Hans Herrmann for initiating this
study at one of the famous coffee seminars at the HLRZ
and Cristian Moukarzel for discussions and for
providing us with Figure~\ref{vrl_1}(b).
K.B.L.\ gratefully acknowledges the support
from the \mbox{Danish} Research Academy and the Carlsberg Foundation.



\nonumsection{References}

\end{document}